\documentclass[PAPER, atlasdraft=false, cernpreprint, USenglish, texmf, orcidlogo]{atlasdoc}

\usepackage{atlaspackage}
\usepackage{atlasbiblatex}
\usepackage{atlasphysics}
\graphicspath{{logos/}{figures/}}

\usepackage{hyperref}

\newcommand{\nch}{N_{\mathrm{ch}}}
\newcommand{\nchr}{N_\mathrm{ch}^\mathrm{rec}}

\newcommand{\snn}{\mbox{$\sqrt{s_{\mathrm{NN}}}$}}
\newcommand{\pTref}{p_{\mathrm{T}}^{\mathrm{ref}}}
\newcommand{\pTbar}{\bar{p}_{\mathrm{T}}}
\newcommand{\pa}{p_{\mathrm{T1}}}
\newcommand{\pb}{p_{\mathrm{T2}}}
\newcommand{\ea}{\eta_{\mathrm{1}}}
\newcommand{\eb}{\eta_{\mathrm{2}}}
\newcommand{\efcal}{E_{\mathrm{T}}^{\mathrm{FCal}}}
\newcommand{\lr}[1]{\left\langle #1\right\rangle}

\newcommand{\HIJING}{\textsc{Hijing}}

\AtlasTitle{Evidence for the collective nature of radial flow in Pb+Pb collisions with the ATLAS detector}
\AtlasAbstract{
Anisotropic flow and radial flow are two key probes of the expansion dynamics and properties of the quark-gluon plasma (QGP). While anisotropic flow has been extensively studied, radial flow, which governs the system’s radial expansion, has received less attention. Notably, direct experimental evidence for the global and collective nature of radial flow fluctuations has been lacking. This Letter presents the first measurement of transverse momentum ($p_{\mathrm{T}}$) dependence of radial flow fluctuations ($v_0(p_{\mathrm{T}})$) over $0.5<p_{\mathrm{T}}<10$ GeV and demonstrates its collective nature using a two-particle correlation method in Pb+Pb collisions at $\sqrt{s_{\mathrm{NN}}}=5.02$ TeV. The data reveal three key features supporting the collective nature of radial flow : long-range correlation in pseudorapidity, factorization in $p_{\mathrm{T}}$, and centrality-independent shape in $p_{\mathrm{T}}$. The comparison with a hydrodynamic model demonstrates the sensitivity of $v_0(p_{\mathrm{T}})$ to bulk viscosity, a crucial transport property of the QGP. These findings establish a new, powerful tool for probing collective dynamics and properties of the QGP.
}

\AtlasRefCode{HION-2024-06}
\PreprintIdNumber{CERN-EP-2025-068}
\AtlasJournal{Physical Review Letters}
\AtlasJournalRef{\PRL 136 032301 (2026)}
\AtlasDOI{https://doi.org/10.1103/ldcn-r2lq}
\hypersetup{pdftitle={ATLAS document},pdfauthor={The ATLAS Collaboration}}

\begin{document}
\maketitle{\textbf{Introduction.}} Heavy-ion collisions at the Large Hadron Collider (LHC) and the Relativistic Heavy Ion Collider (RHIC) provide unique conditions for creating and studying the quark–gluon plasma (QGP). The spacetime evolution of the QGP is well described by relativistic hydrodynamics~\cite{Busza:2018rrf,Heinz:2024jwu}, where the system, driven by pressure gradients, expands collectively and cools until hadronization. This gradient-driven expansion turns geometric anisotropies and size variations in the initial state into two distinct final-state collective phenomena: anisotropic flow ($v_n$, $n= 1,2, \cdots$) and radial flow ($v_0$), respectively. Anisotropic flow quantifies azimuthal momentum anisotropy, while radial flow characterizes the radial boost of the system, which influences the average transverse momentum of particles $[\pT]$ in each event~\cite{Heinz:2013th}.

Extensive measurements of anisotropic flow, particularly differential studies of $v_n(\pT)$ and their event-by-event (EbE) fluctuations~\cite{STAR:2004jwm,PHOBOS:2010ekr,ALICE:2011ab,Adare:2011tg,ATLAS:2012at,Chatrchyan:2013kba,Aad:2013xma,Aad:2014fla,Aad:2015lwa,ATLAS:2019peb}, have provided crucial insights into the initial conditions and transport properties of the QGP. In contrast, experimental studies of radial flow have been more limited. Traditional approaches focus on particle spectra and moments of the $[\pT]$ distribution over event ensembles, such as the mean $\lr{[\pT]}$~\cite{Schnedermann:1993ws} and the scaled variance $v_0 = \sqrt{\lr{(\delta[\pT])^2}}/\lr{[\pT]}$~\cite{NA49:1999inh,ALICE:2014gvd,Adam:2019rsf,ATLAS:2024jvf,STAR:2024wgy}, where $\delta[\pT] = [\pT] - \lr{[\pT]}$. While the $v_0$ observable characterizes global EbE radial flow fluctuations~\cite{Bozek:2012fw,Samanta:2023amp} and scales with the square root of charged-particle multiplicity~\cite{ATLAS:2024jvf}, it provides only the $\pT$-integral information about the radial flow.

Anisotropic and radial flow are thought to originate from single-particle distributions and hence are expected to exhibit long-range correlations in pseudorapidity ($\eta$)~\cite{Heinz:2013th,Jia:2014jca}. For anisotropic flow, these two features have been verified by the approximate factorization of the coefficients of two-particle azimuthal correlation into single-particle coefficients: $V_{n\Delta}(\pa,\pb) \approx v_n(\pa) v_n(\pb)$~\cite{ATLAS:2012at,Jia:2014jca} and $V_{n\Delta}(\ea,\eb) \approx v_n(\ea) v_n(\eb)$~\cite{CMS:2015xmx,ATLAS:2017rij,ATLAS:2020sgl}. Furthermore, $v_n(\pT)$ divided by the integrated flow $v_n$ exhibits a near-universal behavior across centralities~\cite{ATLAS:2018ezv}. This universality arises because the magnitude of $v_n(\pT)$ scales with initial state eccentricity, while its shape, characterized by $k_n(\pT)=v_n(\pT)/v_n$, reflects the hydrodynamic response, which is nearly independent of centrality~\cite{Basar:2013hea}. While this factorization and centrality-independent scaling hold at $\pT\lesssim4$~GeV in central to midcentral Pb+Pb collisions, deviations are observed in peripheral collisions or at higher $\pT$, attributable to non-flow correlations such as resonance decays or jet fragmentation. These direct correlation measurements have established the multi-particle collective nature of anisotropic flow.

In contrast to anisotropic flow, the collective nature of radial flow has traditionally been inferred from model studies of identified particle $\pT$ spectra and their $[\pT]$ and $v_0$~\cite{Hirano:2007ei,Schnedermann:1993ws}. These model-dependent approaches, incorporating radial expansion, have enabled acecss to average radial flow velocities and temperatures~\cite{ALICE:2019hno}. Yet, they do not directly probe the many-body nature of the radial flow, as has been achieved for anisotropic flow. A new observable, $v_0(\pT)$, was proposed recently~\cite{Schenke:2020uqq} to fill this gap. Unlike anisotropic flow, which is more sensitive to shear viscosity, $v_0(\pT)$ is predicted to be more sensitive to bulk viscosity~\cite{Schenke:2020uqq,Parida:2024ckk}, which remains less well constrained. Precision measurements of $v_0(\pT)$, including factorization and scaling checks as well as particle species dependence, thus offer valuable opportunities to constrain the properties of the QGP~\cite{Ryu:2015vwa,JETSCAPE:2020mzn} and possibly disentangle contributions from flow velocity fluctuations and thermal fluctuations~\cite{Retiere:2003kf}.

\begin{figure}[h!]
\centering
\includegraphics[width=0.7\linewidth]{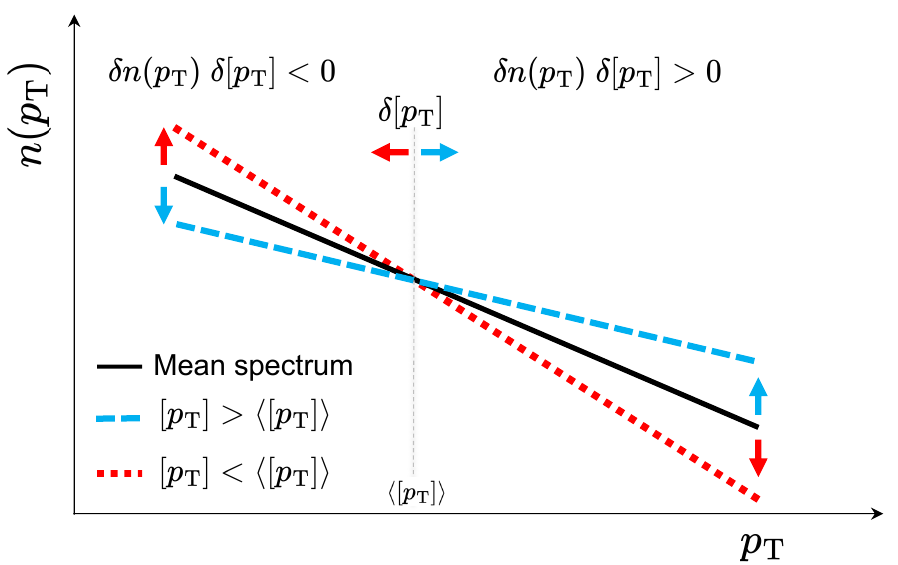}
\caption{Schematic illustration of how radial flow fluctuations lead to correlations between the EbE $\pT$-differential yield $n(\pT)$ and the EbE average transverse momentum $[\pT]$. The blue curve represents an event with larger-than-average radial flow, the red curve an event with smaller-than-average radial flow, and the black curve the ensemble-averaged spectrum.  Events with larger radial flow have flatter spectra, resulting in a positive covariance ($\lr{\delta n(\pT)\delta[\pT]}>0$) at $\pT\gtrsim \lr{[\pT]}$ and a negative covariance ($\lr{\delta n(\pT)\delta[\pT]}<0$) at $\pT\lesssim \lr{[\pT]}$~\cite{Parida:2024ckk}. The zero-crossing point of $v_0(\pT)$, $\pT \approx \langle[\pT] \rangle$, depends on the $\pT$ range used for measurement, which is 0.5--10~GeV in this analysis, but does not depend on $\pTref$ (see Ref.~\cite{Bhatta:2025oyp}).}
\label{fig:1}
\end{figure}

The $v_0(\pT)$ observable is defined based on the correlation between particle spectra in narrow $\pT$ ranges, $N(\pT)$, and $[\pT]$~\cite{Schenke:2020uqq,Parida:2024ckk} (see Figure~\ref{fig:1}). To avoid fluctuations in the total multiplicity and capture only shape variations~\cite{Bhatta:2025oyp}, the fractional spectrum $n(\pT)=N(\pT)/\!\int\!N(\pT)d\pT$ in each event is used. Events with a larger (smaller) radial flow feature a flatter (steeper) distribution of $n(\pT)$ and a correspondingly higher (lower) $[\pT]$. The resulting fluctuations in the spectrum, $\delta n(\pT)  =  n(\pT) - \lr{n(\pT)}$, and in $\delta[\pT]$ are correlated in a characteristic pattern: negative correlation at $\pT \lesssim \lr{[\pT]}$ and positive at $\pT \gtrsim \lr{[\pT]}$. By assuming a factorization of the normalized covariance~\cite{Schenke:2020uqq,Parida:2024ckk}:
\begin{align}\label{eq:1}
\frac{\lr{\delta n(\pT) \delta[\pT]}}{\lr{n(\pT)}\lr{[\pT]}} = v_0(\pT) v_0\;,
\end{align}
$v_0(\pT)$ can be calculated from the data. The normalization reduces sensitivity to the particle reconstruction efficiency. If radial flow is collective, $v_0$ is related to $v_0(\pT)$ by an approximate sum rule~\cite{Schenke:2020uqq,Bhatta:2025oyp}:
\small{\begin{align}\label{eq:2}
v_0\! \equiv\!  \frac{\sqrt{\lr{(\delta[\!\pT\!]\!)^2}}_{\!A}}{\lr{[\pT]}_{A}} \!\approx\!  \frac{\int_{\pT\in A}\! (\pT\!-\!\lr{[\!\pT\!]}_{\!A})\! \lr{n(\!\pT\!)}\! v_0(\pT\!) d\pT}{\int_{\pT\in A} \pT \lr{n(\pT)} d\pT},
\end{align}}\normalsize
where ``$A$'' denotes the reference-$\pT$ ($\pTref$) range for integration and calculating $v_0$. The measured $v_0(\pT)$ is expected to be independent of $\pTref$ and exhibit centrality-independent scaling. However, this factorization and scaling can be broken in the presence of significant non-flow correlations.

This Letter presents measurements of $v_0(\pT)$ using 470~$\mu\mathrm{b}^{-1}$ of Pb+Pb collision data at $\snn=5.02$~TeV from 2015. Results are obtained for several $\pTref$ and various centrality ranges.  Comparison with a hydrodynamic model further reveals the impact of QGP viscosities, particularly bulk viscosity, on radial flow fluctuations. This analysis closely follows a previous ATLAS analysis that measured $v_0$ and its centrality scaling properties~\cite{ATLAS:2024jvf}.

\textbf{ATLAS detector and data selection.} The measurements are performed using the ATLAS inner detector (ID), forward calorimeter (FCal), and zero-degree calorimeters (ZDCs) along with the trigger and data acquisition systems~\cite{ATLAS:2008xda, ATLAS:2010ylv,ATLAS:2016wtr}. The ID detects charged particles within $|\eta|<2.5$~\footnote{ATLAS uses a right-handed coordinate system with its origin at the nominal interaction point in the center of the detector and the z-axis along the beam pipe. The x-axis points from the IP to the center of the LHC ring, and the y-axis points upward. Cylindrical coordinates (r,$\phi$) are used in the transverse plane, $\phi$ being the azimuthal angle around the beam pipe. The pseudorapidity is defined in terms of the polar angle $\theta$ as $\eta$=-$\ln{\tan{\theta/2}}$.} using a combination of silicon pixel and microstrip detectors, along with a straw-tube transition-radiation tracker, all immersed in a 2 T axial magnetic field~\cite{ATLAS:2010ylv}. The FCal consists of three sampling layers, covering $3.2<|\eta|<4.9$. The ZDCs are positioned at $\pm 140 \mathrm{~m}$ from the interaction point and detect neutrons with $|\eta|>8.3$. The ATLAS trigger system~\cite{ATLAS:2016wtr} consists of a hardware-based level-one (L1) and software-based high-level triggers. A software suite~\cite{ATLAS:2024vdo} is used in data simulation, in the reconstruction and analysis of real and simulated data, in the detector operations, and in the trigger and data acquisition systems.

Pb+Pb events are selected by requiring the total transverse energy deposited in calorimeters over $|\eta|<4.9$ at L1 to exceed 50~GeV. Additionally, dedicated central collision triggers enhance the number of events with large FCal transverse energy~\cite{ATLAS:2019peb}.

Charged-particle tracks are reconstructed from hits in the ID using a procedure optimized for heavy-ion collisions~\cite{ATLAS:2015qmb}. The distance of the closest approach of the track to the primary vertex must be within 1.25 mm in both the transverse and longitudinal directions. More details of the track selection are found in Ref.~\cite{ATLAS:2018ezv}. Tracks used in this analysis have $0.5<\pT<10$~GeV and $|\eta|<2.5$. This $\pT$ range spans the regime predominantly influenced by radial flow as well as the transition region where jet-quenching effects become significant. The total number of such tracks is denoted by $\nchr$. Events containing multiple inelastic collisions (pileup) are suppressed by exploiting the correlation between $\nchr$ and the transverse energy measured in the FCal, $\efcal$, as well as between the energy deposited in the ZDCs and $\efcal$. The residual pileup probability is 0.17\% among all selected events, but is smaller than 0.01\% in central collisions~\cite{ATLAS:2022kqu,ATLAS:2024jvf}.

Events are categorized into centrality ranges using a parametrization of the $\efcal$ distribution \cite{ATLAS:2019peb} based on a Glauber model \cite{Miller:2007ri}, starting at 0\% for the most central collisions with the highest $\efcal$ value and ending at 70\%. Events in this centrality range are fully efficient against the L1 event selection criteria.

\textbf{Analysis.}  The track reconstruction efficiency, $\epsilon\left(\pT, \eta, \nchr\right)$, is evaluated using Monte Carlo (MC) simulated events for Pb+Pb collisions generated with $\HIJING$~\cite{Gyulassy:1994ew} and charged particles definition according to Ref.~\cite{ATLAS:2015qmb}. The detector response is simulated using \GEANT~\cite{GEANT4:2002zbu,ATLAS:2010arf}, and events are reconstructed with the same algorithms as used for the data. The HIJING spectra have been reweighted to match the data. In central Pb+Pb collisions, the efficiency ranges from 71\% at $\eta \approx 0$ to about 40\% for $|\eta|>2$ at $\pT>0.8$~GeV, and it decreases by $12\%$ from 0.8 to 0.5~GeV in $\pT$. In peripheral collisions, the efficiency is up to 4\% higher. The rate of falsely reconstructed (``fake'') tracks, $f\left(\pT, \eta, \nchr\right)$, is significant for $\pT<1$~GeV in central collisions, ranging from 2\% for $|\eta|<1$ to 8\% at larger $|\eta|$. The fake-track rate drops rapidly for higher $\pT$ and more peripheral collisions.

The observables in this analysis are obtained using a two-subevent method~\cite{Jia:2017hbm,Bhatta:2021qfk}, which takes particles from two $\eta$ regions labeled $a$ and $b$: $-2.5<\eta_a<-\frac{\eta_{\mathrm{gap}}}{2}$ and $\frac{\eta_{\mathrm{gap}}}{2}<\eta_b<2.5$. This subevent method was first used to measure the $v_0$ at RHIC in Ref.~\cite{STAR:2024wgy}. Four choices of gaps, $\eta_{\mathrm{gap}}=0,1,2$, and 3, are used to evaluate the effects of short-range non-flow correlations as well as possible longitudinal radial flow decorrelations~\cite{Chatterjee:2017mhc}. Each reconstructed charged track $i$ is assigned a weight $w_i$ to correct for reconstruction efficiency $\epsilon_i$ and fake rate $f_i$ : $w_i \equiv\left(1-f_i\right) / \epsilon_i$ \cite{ATLAS:2019pvn}. The charged-particle multiplicity, corrected for detector effects, in $0.5 <\pT < 5$~GeV and $|\eta| < 2.5$ is obtained as $\nch=\Sigma_i w_i$.

For each collision subevent, the following quantities are calculated: $\left[\pT\right]=\Sigma_i w_i p_{\mathrm{T}, i} /\Sigma_i w_i$, $\delta[\pT]=\Sigma_{i} w_i \delta p_{\mathrm{T}, i} / \Sigma_{i} w_i$ with $\delta p_{\mathrm{T}, i} \equiv$ $p_{\mathrm{T}, i}-\left\langle\left[\pT\right]\right\rangle$, and $n(\pT) = \Sigma_i^{\pT} w_i /\Sigma_i w_i$, where $\Sigma_i^{\pT}$ represents the sum over all tracks in a narrow $\pT$ range. The two-particle observables are calculated in narrow $\nch$ ranges as: $\lr{(\delta[\pT])^2}= \lr{\delta[\pT]_a\delta[\pT]_b}$, \mbox{$\lr{\delta n(\pT) \delta[\pT]}\! =\!\frac{1}{2}\!\lr{\delta n_a(\pT) \delta[\pT]_b\!+\!\delta n_b(\pT) \delta[\pT]_a}$}. The measurements are performed for three $\pTref$ ranges, 0.5--2, 0.5--5, and 1--5~GeV; according to Eq.~\eqref{eq:1}, $v_0$ can only be calculated in fairly wide $\pTref$ ranges, motivating the partial overlap in their choices. Statistical uncertainties for all observables are computed using a standard Poisson bootstrap method.

Systematic uncertainties stem from track selection, reconstruction efficiency, residual pileup, centrality definition, and MC closure test of the analysis procedure. These uncertainties obtained in $0.5<\pTref<2$~GeV and $\eta_{\mathrm{gap}}=1$ are summarized below for $v_0$ and for $v_0(\pT)$ at a low $\pT$ (0.5--0.8~GeV) and a high $\pT$ region (2--3~GeV).

Track selection uncertainties are assessed by comparing nominal results with those obtained under looser or stricter criteria. This results in a deviation of 0.5--5\% for $v_0$, and 1--8\% at low $\pT$ and 0.5--1.5\% at high $\pT$ for $v_0(\pT)$. The $v_0(\pT)$ as defined in Eq.~\eqref{eq:1} is insensitive to the track reconstruction efficiency. Indeed, repeating the analysis without track weights leads to variations in $v_0(\pT)$ smaller than uncertainties due to track selection. Uncertainties in the efficiency, mainly due to the modeling of the detector material in \GEANT, can be as large as 4\%. Their impacts are within 0.5--1\% for $v_0$ and 2--3\% at low $\pT$ and 0.5--1\% at high $\pT$ for $v_0(\pT)$. The impact of the azimuthal efficiency variation is assessed by including it in the track weight; no observable changes are found in the results. Residual pileup effects are estimated by adjusting the pileup rejection criteria, leading to uncertainties less than 0.5\% for all observables. Centrality definition uncertainties, assessed by varying the Glauber model parameters, are relevant only when results are presented in centrality ranges and are below 3\% for all observables.

The $\HIJING$ MC samples are used to evaluate the consistency of calculated observables, obtained using truth particles or the reconstructed tracks with the same correction procedures as for the real data applied~\cite{ATLAS:2019peb}. This check also accounts for the smearing of measured $\left[\pT\right]$ and $n(\pT)$ from their true values. The differences are about 1\% for $v_0$ and 2\% for $v_0(\pT)$.

The full systematic uncertainties for each observable are obtained by adding contributions from individual sources in quadrature. Track selection is the dominant source of systematic uncertainty, particularly in midcentral and central collisions. The total uncertainties are 1.5--5\% for $v_0$, and 3--8.5\% for $v_0(\pT)$ at low $\pT$ and 2--3.5\% at high $\pT$. The uncertainties are largest in central collisions and reduce to less than half their size in peripheral collisions. The uncertainties are generally smaller for the other two $\pTref$ ranges, and are somewhat larger for bigger $\eta_{\mathrm{gap}}$. The total uncertainties for each data point are obtained by combining statistical and systematic uncertainties, which are then shown alongside the statistical uncertainties when presenting the results.

\begin{figure}[h!]
\centering
\includegraphics[width=0.8\linewidth]{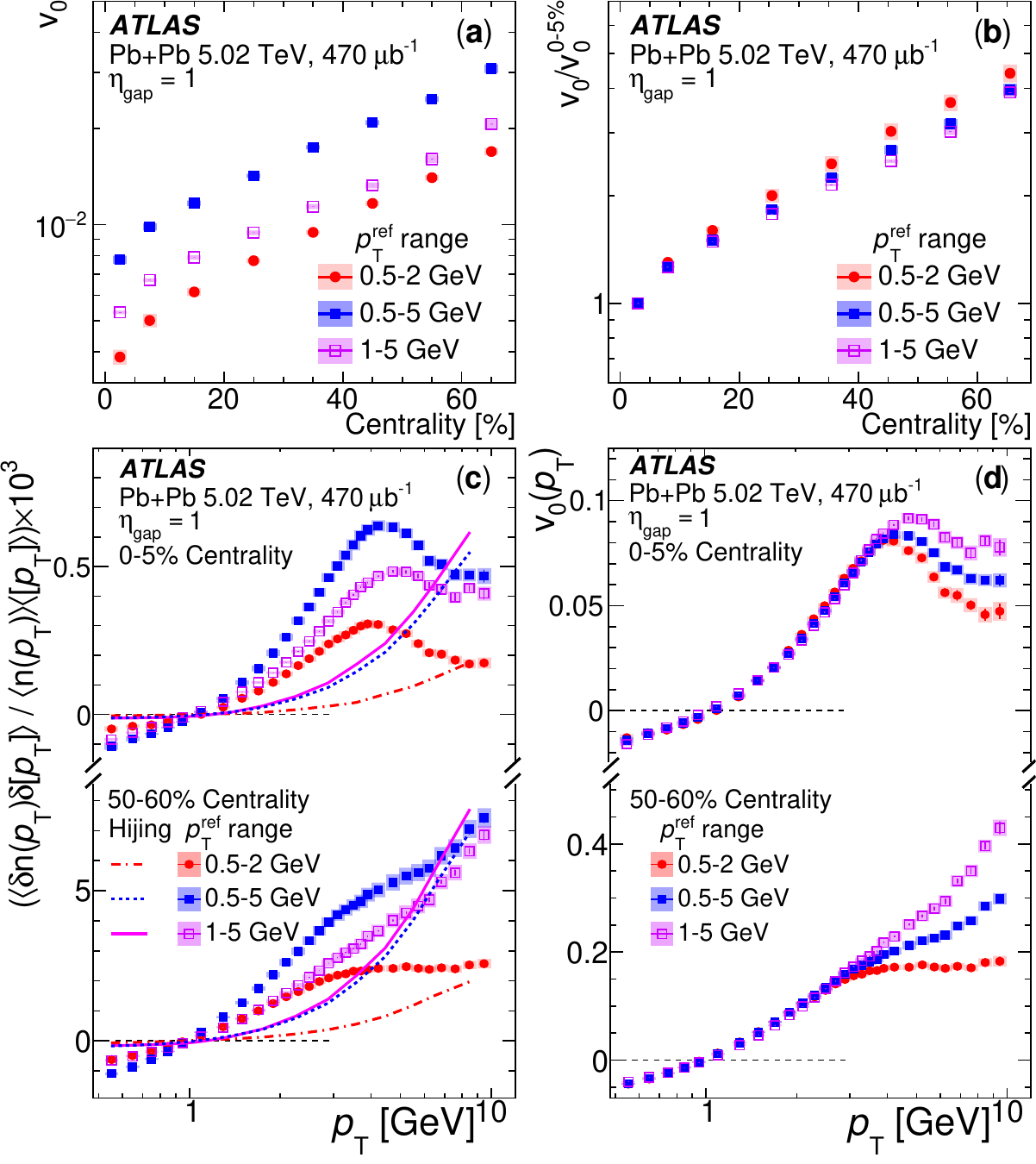}
\caption{Centrality dependence of $v_0$ (\textbf{a}) and $v_0/v_0^{0-5\%}$ (\textbf{b}) for three $\pTref$ ranges, and normalized covariance from Eq.~\ref{eq:1} (\textbf{c}) and $v_0(\pT)$ (\textbf{d}) obtained in 0--5\% centrality (top) and 50--60\% centrality (bottom) in three $\pTref$ ranges. The color lines indicate the predictions from the $\HIJING$ model containing only non-flow correlations. Bars and shaded areas indicate statistical and total uncertainties, respectively.}
\label{fig:2}
\end{figure}

\textbf{Results.} Figure~\ref{fig:2}a shows the centrality dependence of $v_0$ computed in three $\pTref$ ranges. The magnitude of $v_0$ depends on the choice of $\pTref$, and shows a characteristic decrease toward central collisions as observed in Ref.~\cite{ATLAS:2024jvf}. It is the largest for the widest $\pTref$ range, as expected from Eq.~\eqref{eq:2}. However, when dividing $v_0$ by its value in the 0--5\% centrality range, $v_0/v_0^{0-5\%}$, the trend in centrality is similar across all $\pTref$ selections, as shown in Figure~\ref{fig:2}b. Hence, $v_0$ captures EbE fluctuations in the overall spectral shape, consistent with a collective nature of radial flow fluctuations.

To quantify this collective nature, the factorization of the covariance into a product of $v_0(\pT)$ and $v_0$ from Eq.~\eqref{eq:1} is examined in Figures~\ref{fig:2}c and \ref{fig:2}d for two representative centralities (0--5\% and 50--60\%). While the normalized covariance varies substantially with $\pTref$, the extracted $v_0(\pT)$ remains nearly $\pTref$-independent up to $\pT \approx 3$~GeV. This factorization, akin to what is observed for anisotropic flow, further supports the collective nature of radial flow. The $v_0(\pT)$ data for the three $\pTref$ cross zero at $\pT\approx 1$~GeV, approximately the $\lr{[\pT]}$ for the 0.5--10~GeV range. The $v_0(\pT)$ data are also found to satisfy the sum rule in Eq.~\eqref{eq:2} within 4\%. Above $\pT \approx 3$~GeV and in more peripheral collisions, the data become dependent on $\pTref$ due to larger non-flow contributions. Among the three reference ranges, results for $\pTref \in 0.5$--2~GeV are least affected by non-flow correlations, as suggested by the trend of $v_0(\pT)$ at high $\pT$.

In central collisions, $v_0(\pT)$ shows a decrease above $\pT\approx 4$~GeV, similar to anisotropic flow, which might be associated with a transition from a flow-dominated regime to a jet-quenching dominated regime~\cite{Schenke:2020uqq}, which changes the shape of the underlying spectrum and its EbE fluctuations~\cite{Jia:2025rab}. No such decrease is observed in peripheral collisions.

In Figure~\ref{fig:2}c, the covariance is compared to calculations from the $\HIJING$ model containing only non-flow correlations. $\HIJING$ underpredicts the magnitude of the data at $\pT<6$~GeV, implying that the collective radial flow generates most of the signal in the data. Above $\pT\approx 6$~GeV, $\HIJING$ predictions become comparable to the data, suggesting that the high $\pT$ behavior of the covariance could be impacted by non-flow correlations, especially in peripheral collisions.

\begin{figure}[h!]
\centering
\includegraphics[width=0.8\linewidth]{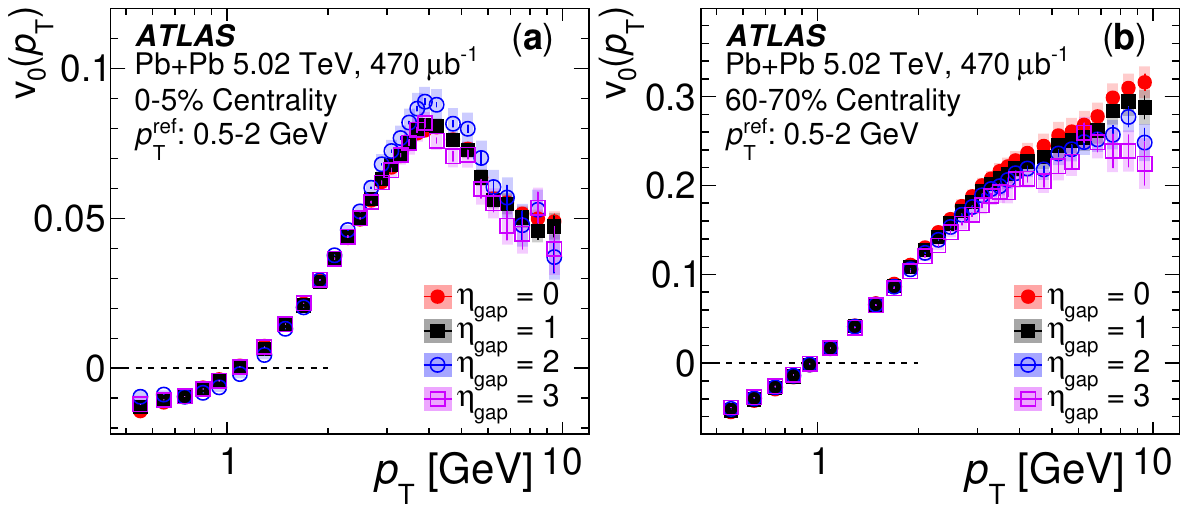}
\caption{$v_0(\pT)$ for $\eta_{\mathrm{gap}}$=0, 1, 2, and 3 in 0--5\%  (\textbf{a}) and 60--70\%  (\textbf{b}) centrality. Bars and shaded areas indicate statistical and total uncertainties, respectively.}
\label{fig:3}
\end{figure}

The impact of non-flow correlations on $v_0(\pT)$ is further investigated by varying $\eta_{\mathrm{gap}}$ between the subevents. A larger $\eta_{\mathrm{gap}}$ reduces the short-range non-flow contributions but also enhances the flow decorrelation effects~\cite{Chatterjee:2017mhc,ATLAS:2020sgl}. Figure~\ref{fig:3} shows the $v_0(\pT)$ data in 0--5\% and 60--70\% centrality ranges for different $\eta_{\mathrm{gap}}$ selections. Little change is observed in central collisions when $\eta_{\mathrm{gap}}$ is varied from 0 to 3. In contrast, $v_0(\pT)$ in peripheral collisions at $\pT > 3$~GeV decreases slightly, which is consistent with residual non-flow contributions from away-side jet fragmentation. The weak dependence on $\eta_{\mathrm{gap}}$ supports the interpretation that radial flow is a long-range, global correlation similar to anisotropic flow. Based on these observations, $\eta_{\mathrm{gap}}=1$ is used as the default configuration.

\begin{figure}[h!]
\centering
\includegraphics[width=0.8\linewidth]{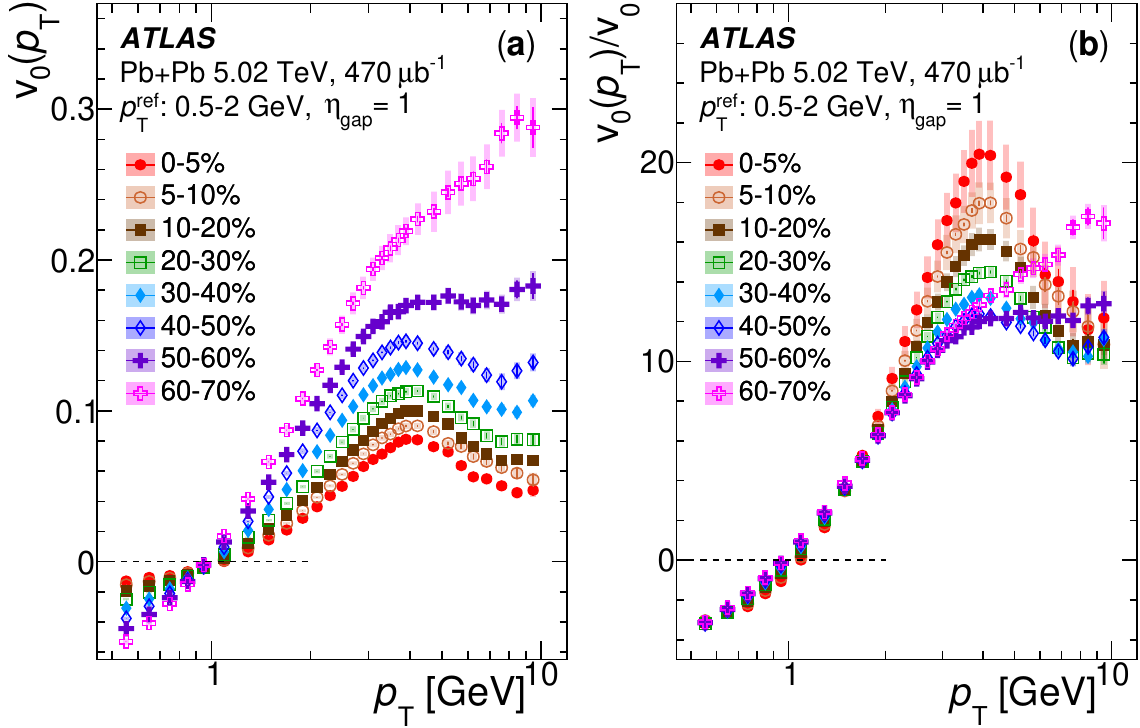}
\caption{The $v_0(\pT)$ (\textbf{a}) and $v_0(\pT)/v_0$  (\textbf{b}) for different centrality ranges. Bars and shaded areas represent statistical and total uncertainties, respectively.}
\label{fig:4}
\end{figure}

Next, the shape of $v_0(\pT)$ across centralities is investigated using the normalized quantity $v_0(\pT)/v_0$, which has the advantage of being insensitive to the initial-state effects~\cite{Schenke:2020uqq}. Figure~\ref{fig:4} compares $v_0(\pT)$ and $v_0(\pT)/v_0$ for several centrality ranges. While $v_0(\pT)$ exhibits a pronounced centrality dependence, $v_0(\pT)/v_0$ is almost independent of centrality in the range $\pT\lesssim 2.5$~GeV. This observation indicates that the magnitude of $v_0(\pT)$ is primarily driven by the fluctuations in nuclear overlap area and density, which is larger in peripheral collisions~\cite{Bozek:2012fw,ATLAS:2024jvf}. In contrast, the shape of $v_0(\pT)$ is governed by the hydrodynamic response $k_0(\pT)\approx v_0(\pT)/v_0$~\cite{Parida:2024ckk}, which is nearly independent of centrality. This scaling behavior is analogous to that observed for $v_n(\pT)/v_n$. However, $v_n(\pT)/v_n$ is sensitive to both shear and bulk viscosities, while $v_0(\pT)/v_0$ is mostly sensitive to bulk viscosity~\cite{Heinz:2013th}.

Above $\pT\approx 3$~GeV, where the non-flow contribution is more significant, this scaling is imperfect. Up to 40--50\% centrality, $v_0(\pT)/v_0$ is observed to first rise and then fall, which could be due to the change in the baryon-meson composition or onset of jet quenching that reduces the fractional momentum fluctuations~\cite{Jia:2025rab}. For more peripheral collisions, the data exhibit either a flat behavior or an increasing trend with $\pT$, consistent with a larger non-flow contribution at high $\pT$ and in peripheral collisions.

\begin{figure}[h!]
\centering
\includegraphics[width=0.8\linewidth]{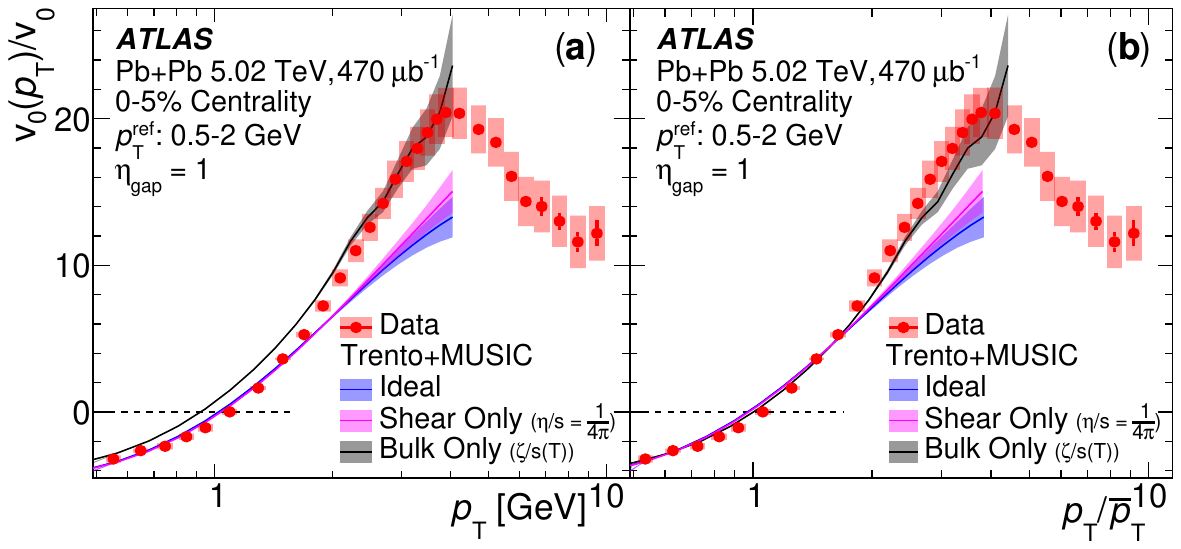}
\caption{The $v_0(\pT)/v_0$ as a function of $\pT$ (\textbf{a}) and $\pT/\pTbar$ (\textbf{b}) in 0--5\% centrality range, where $\pTbar$ is the mean $\pT$ in 0.5--10~GeV. Bars and shaded areas represent statistical and total uncertainties, respectively. They are compared with hydrodynamic model predictions without viscosity (blue), with a constant shear viscosity (in terms of shear viscosity to entropy density ratio of $\eta/s = \frac{1}{4\pi}$, magenta), and with a temperature-dependent bulk viscosity (grey)~\cite{Ryu:2015vwa,Ryu:2017qzn}. The shaded bands on the model curves represent the statistical uncertainties.}
\label{fig:5}
\end{figure}

Figure~\ref{fig:5} compares the data from the 0--5\% centrality range to hydrodynamic model predictions from Ref.~\cite{Parida:2024ckk}, based on the Trento+MUSIC framework with either constant shear viscosity or temperature-dependent bulk viscosity from Refs.~\cite{Ryu:2015vwa,Ryu:2017qzn}. Since these calculations are not fully tuned to reproduce overall $\pT$ spectra and fluctuations, following conformal scaling arguments~\cite{Basar:2013hea}, the authors recommend comparing $v_0(\pT)/v_0$ as a function of $\pT/\pTbar$,  where $\pTbar$ is the average transverse momentum in the range of 0.5--10~GeV in this analysis. The ratio $v_0(\pT)/v_0$ reduces the dependence on initial conditions, while the scaled quantity $\pT/\pTbar$ reduces dependencies on model descriptions of $[\pT]$.

As shown in Figure~\ref{fig:5}, calculations with and without shear viscosity lead to similar predictions, whereas calculation with bulk viscosity alters both the zero-crossing in $\pT$ and the high-$\pT$ behavior. Without bulk viscosity, the model underestimates the data at high $\pT$, while using a standard temperature-dependent bulk viscosity~\cite{Ryu:2015vwa,Ryu:2017qzn} improves agreement with the data, albeit reducing the zero-crossing point. The zero-crossing point shift reflects a reduction of average transverse momentum in the presence of bulk viscosity. This reduction is expected to be systematic across centralities and should preserve the universal scaling observed in Fig.~\ref{fig:4}b. Rescaling the horizontal axis by $\pTbar$ further improves the agreement, isolating the clear sensitivity of the $v_0(\pT)$ data at $\pT> 2$ GeV to the bulk viscosity of the QGP. Further model tuning for a precise description of the initial conditions and other bulk observables, such as yields of baryons and mesons and $\nch$ distribution, is expected to improve the description of the data and provide more precise constraints on the bulk viscosity. Our measurement provides the foundation for such quantitative constraints.

\FloatBarrier
\textbf{Summary.} This Letter reports the first measurement of the $\pT$-differential radial flow fluctuations $v_0(\pT)$ using charged particle tracks over the range $0.5<\pT<10$~GeV in Pb+Pb collisions at $\snn$ = 5.02~TeV using the ATLAS detector at the LHC. This observable is extracted from the correlations between each event's $\pT$-differential spectrum and its average transverse momentum $[\pT]$ in separate subevents. A characteristic rise of $v_0(\pT)$ up to 3--4~GeV, a decrease at higher $\pT$, and a sign change near 1~GeV are observed. This rise and fall behavior could be associated with an interplay between collective flow, changes in baryon-meson composition, and jet quenching, although a quantitative understanding of such interplay requires a realistic model calculation. The $v_0(\pT)$ is nearly independent of the pseudorapidity gap between subevents, supporting the long-range nature of the radial flow. These results are confirmed by a similar study from the ALICE Collaboration~\cite{ALICE:2025iud} that appeared after submission of this Letter for publication. A factorization test shows that $v_0(\pT)$ for $\pT\lesssim 3$~GeV is nearly independent of the $\pT$ range used to calculate $[\pT]$, establishing the collective nature of the radial flow. Notably, the shape of $v_0(\pT)$ for $\pT\lesssim 3$~GeV is nearly centrality-independent, implying a universal hydrodynamic response to radial expansion. Taken together, these results confirm three hallmarks of collectivity -- long-range correlation, factorization, and centrality-independent shape -- now demonstrated for radial flow. Comparisons with a hydrodynamic model reveal that $v_0(\pT)$ is sensitive to bulk viscosity, highlighting its potential for quantitative constraints on QGP transport properties.

\section*{Acknowledgements}

%
%

%
%

We thank CERN for the very successful operation of the LHC and its injectors, as well as the support staff at
CERN and at our institutions worldwide without whom ATLAS could not be operated efficiently.

The crucial computing support from all WLCG partners is acknowledged gratefully, in particular from CERN, the ATLAS Tier-1 facilities at TRIUMF/SFU (Canada), NDGF (Denmark, Norway, Sweden), CC-IN2P3 (France), KIT/GridKA (Germany), INFN-CNAF (Italy), NL-T1 (Netherlands), PIC (Spain), RAL (UK) and BNL (USA), the Tier-2 facilities worldwide and large non-WLCG resource providers. Major contributors of computing resources are listed in Ref.~\cite{ATL-SOFT-PUB-2025-001}.

We gratefully acknowledge the support of ANPCyT, Argentina; YerPhI, Armenia; ARC, Australia; BMWFW and FWF, Austria; ANAS, Azerbaijan; CNPq and FAPESP, Brazil; NSERC, NRC and CFI, Canada; CERN; ANID, Chile; CAS, MOST and NSFC, China; Minciencias, Colombia; MEYS CR, Czech Republic; DNRF and DNSRC, Denmark; IN2P3-CNRS and CEA-DRF/IRFU, France; SRNSFG, Georgia; BMFTR, HGF and MPG, Germany; GSRI, Greece; RGC and Hong Kong SAR, China; ICHEP and Academy of Sciences and Humanities, Israel; INFN, Italy; MEXT and JSPS, Japan; CNRST, Morocco; NWO, Netherlands; RCN, Norway; MNiSW, Poland; FCT, Portugal; MNE/IFA, Romania; MSTDI, Serbia; MSSR, Slovakia; ARIS and MVZI, Slovenia; DSI/NRF, South Africa; MICIU/AEI, Spain; SRC and Wallenberg Foundation, Sweden; SERI, SNSF and Cantons of Bern and Geneva, Switzerland; NSTC, Taipei; TENMAK, T\"urkiye; STFC/UKRI, United Kingdom; DOE and NSF, United States of America.

Individual groups and members have received support from BCKDF, CANARIE, CRC and DRAC, Canada; CERN-CZ, FORTE and PRIMUS, Czech Republic; COST, ERC, ERDF, Horizon 2020, ICSC-NextGenerationEU and Marie Sk{\l}odowska-Curie Actions, European Union; Investissements d'Avenir Labex, Investissements d'Avenir Idex and ANR, France; DFG and AvH Foundation, Germany; Herakleitos, Thales and Aristeia programmes co-financed by EU-ESF and the Greek NSRF, Greece; BSF-NSF and MINERVA, Israel; NCN and NAWA, Poland; La Caixa Banking Foundation, CERCA Programme Generalitat de Catalunya and PROMETEO and GenT Programmes Generalitat Valenciana, Spain; G\"{o}ran Gustafssons Stiftelse, Sweden; The Royal Society and Leverhulme Trust, United Kingdom.

In addition, individual members wish to acknowledge support from Armenia: Yerevan Physics Institute (FAPERJ); CERN: European Organization for Nuclear Research (CERN DOCT); Chile: Agencia Nacional de Investigaci\'on y Desarrollo (FONDECYT 1230812, FONDECYT 1240864, Fondecyt 3240661, Fondecyt Regular 1240721); China: Chinese Ministry of Science and Technology (MOST-2023YFA1605700, MOST-2023YFA1609300), National Natural Science Foundation of China (NSFC - 12175119, NSFC 12275265); Czech Republic: Czech Science Foundation (GACR - 24-11373S), Ministry of Education Youth and Sports (ERC-CZ-LL2327, FORTE CZ.02.01.01/00/22\_008/0004632), PRIMUS Research Programme (PRIMUS/21/SCI/017); EU: H2020 European Research Council (ERC - 101002463); European Union: European Research Council (BARD No. 101116429, ERC - 948254, ERC 101089007), European Regional Development Fund (SMASH COFUND 101081355, SLO ERDF), Horizon 2020 Framework Programme (MUCCA - CHIST-ERA-19-XAI-00), European Union, Future Artificial Intelligence Research (FAIR-NextGenerationEU PE00000013), Horizon 2020 (EuroHPC - EHPC-DEV-2024D11-051), Italian Center for High Performance Computing, Big Data and Quantum Computing (ICSC, NextGenerationEU); France: Agence Nationale de la Recherche (ANR-21-CE31-0022, ANR-22-EDIR-0002, ANR-24-CE31-0504-01); Germany: Baden-Württemberg Stiftung (BW Stiftung-Postdoc Eliteprogramme), Deutsche Forschungsgemeinschaft (DFG - 469666862, DFG - CR 312/5-2); China: Research Grants Council (GRF); Italy: Istituto Nazionale di Fisica Nucleare (ICSC, NextGenerationEU), Ministero dell'Università e della Ricerca (NextGenEU 153D23001490006 M4C2.1.1, NextGenEU I53D23000820006 M4C2.1.1, NextGenEU I53D23001490006 M4C2.1.1, SOE2024\_0000023); Japan: Japan Society for the Promotion of Science (JSPS KAKENHI JP22H01227, JSPS KAKENHI JP22H04944, JSPS KAKENHI JP22KK0227, JSPS KAKENHI JP23KK0245, JSPS KAKENHI JP24K23939); Norway: Research Council of Norway (RCN-314472); Poland: Ministry of Science and Higher Education (IDUB AGH, POB8, D4 no 9722), Polish National Science Centre (NCN 2021/42/E/ST2/00350, NCN OPUS 2023/51/B/ST2/02507, NCN OPUS nr 2022/47/B/ST2/03059, NCN UMO-2019/34/E/ST2/00393, UMO-2022/47/O/ST2/00148, UMO-2023/49/B/ST2/04085, UMO-2023/51/B/ST2/00920, UMO-2024/53/N/ST2/00869); Portugal: Foundation for Science and Technology (FCT); Spain: Ministry of Science and Innovation (MCIN \& NextGenEU PCI2022-135018-2, MICIN \& FEDER PID2021-125273NB, RYC2019-028510-I, RYC2020-030254-I, RYC2021-031273-I, RYC2022-038164-I); Sweden: Carl Trygger Foundation (Carl Trygger Foundation CTS 22:2312), Swedish Research Council (Swedish Research Council 2023-04654, VR 2021-03651, VR 2022-03845, VR 2022-04683, VR 2023-03403, VR 2024-05451), Knut and Alice Wallenberg Foundation (KAW 2018.0458, KAW 2022.0358, KAW 2023.0366); Switzerland: Swiss National Science Foundation (SNSF - PCEFP2\_194658); United Kingdom: Leverhulme Trust (Leverhulme Trust RPG-2020-004), Royal Society (NIF-R1-231091); United States of America: U.S. Department of Energy (ECA DE-AC02-76SF00515), Neubauer Family Foundation.

%
%


\section*{Appendix} 
\begin{figure}[hbp!]
\centering
\includegraphics[width=0.64\linewidth]{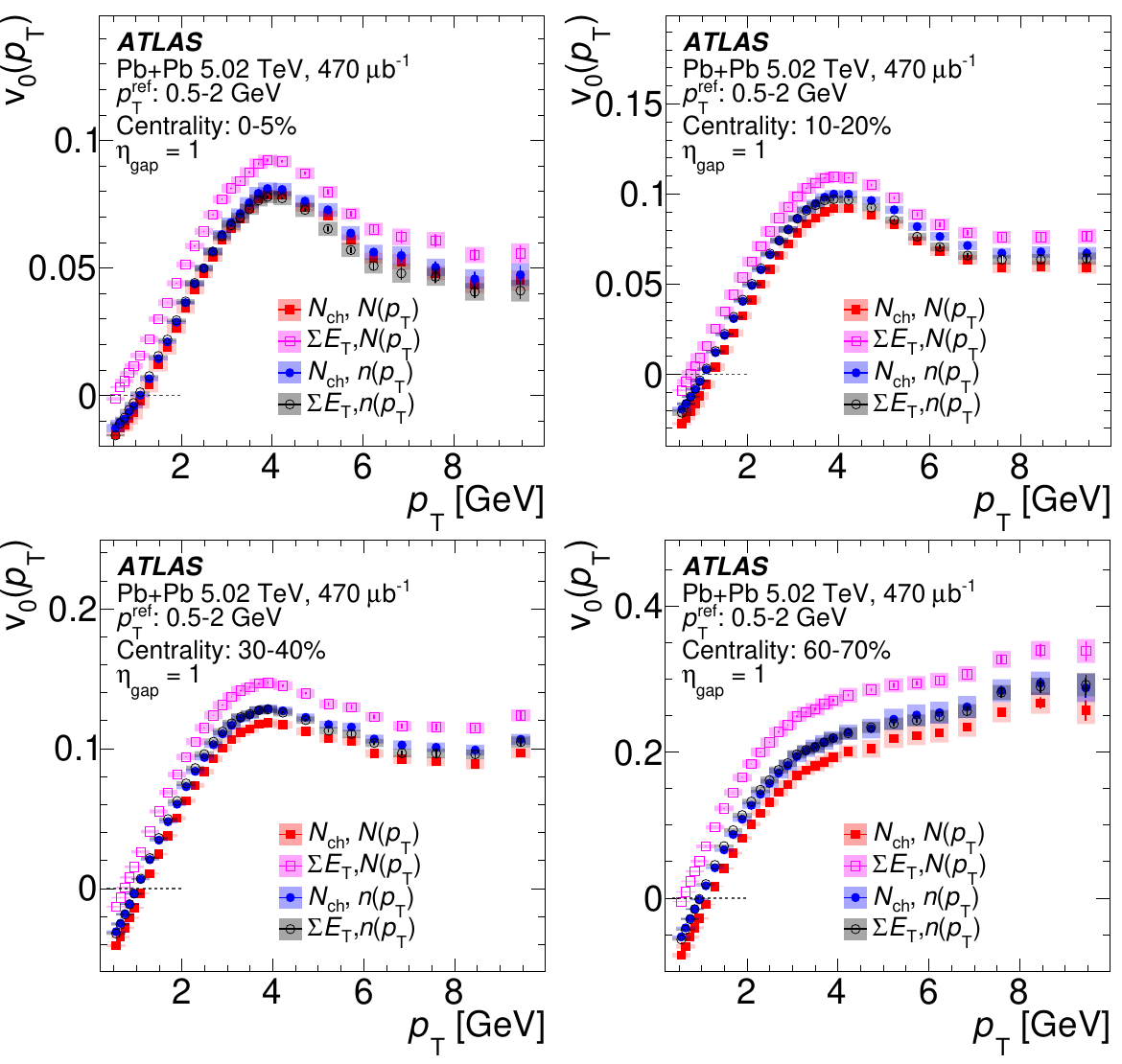}
\vspace*{-0.2cm}\caption{Comparison of $v_0(\pT)$ measured using EbE spectra $N(\pT)$ and EbE normalized spectra $n(\pT)$ for several centralities. The event averages are done by combining observables calculated in narrow slices of $\nch$ or $\efcal$. They are obtained for $\pTref\in$ 0.5--2~GeV and $\eta_{\mathrm{gap}}=1$. Bars and shaded areas indicate statistical and total uncertainties, respectively.}
\label{fig:app1}
\end{figure}

\vspace*{-0.2cm}
The default analysis was performed using normalized EbE spectra, i.e., $n(\pT)$, and then averaging in narrow $\nch$. The impact of residual multiplicity fluctuation in this observable can be checked by repeating the analysis either by using the spectra without normalization and/or performing an average in narrow $\efcal$. This procedure results in four measurements of the $v_0(\pT)$, which are displayed in Figure~\ref{fig:app1} for four centrality ranges. The results show that the global multiplicity fluctuations could shift the $v_0(\pT)$ vertically and change the zero-crossing point in $\pT$, but not the shape of $v_0(\pT)$~\cite{Bhatta:2025oyp}.

\begin{figure}[h!]
\centering
\vspace*{-0.2cm}\includegraphics[width=0.5\linewidth]{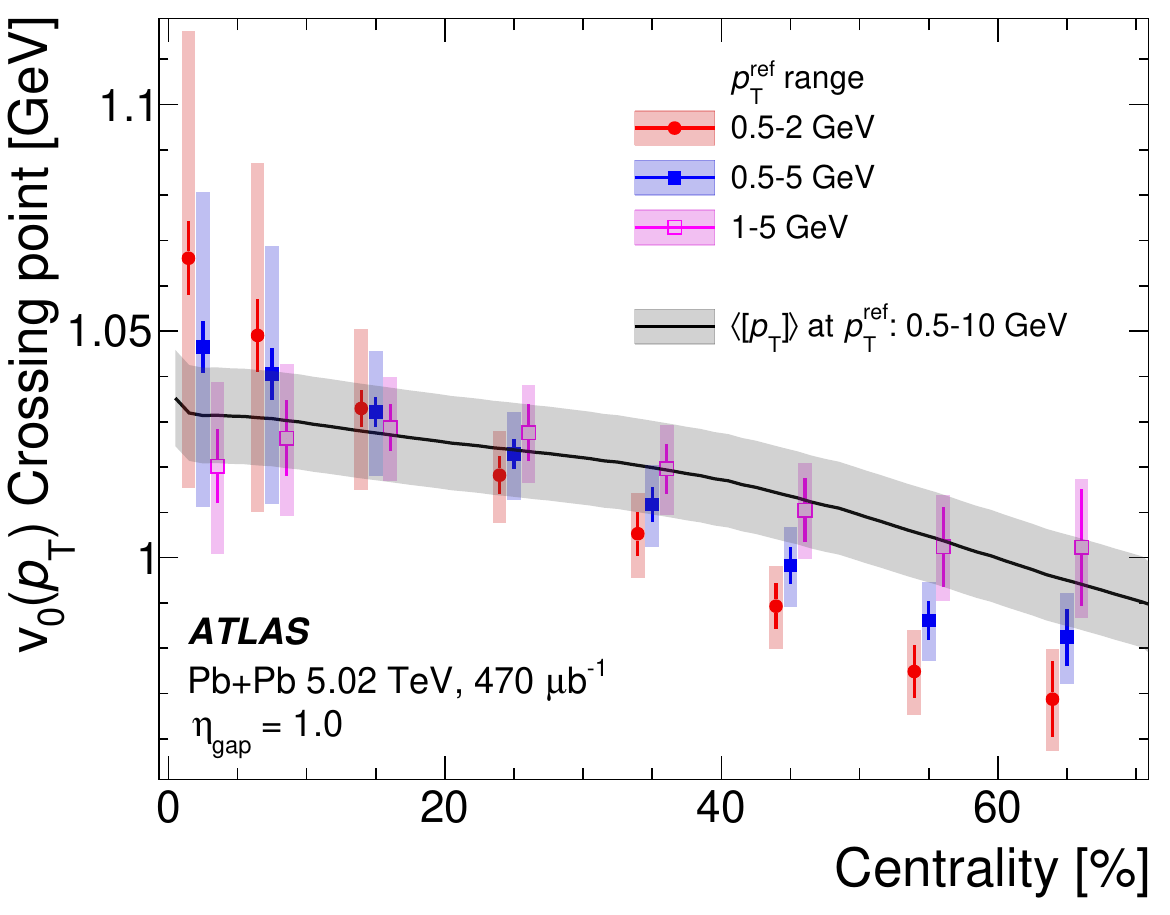}\vspace*{-0.2cm}
\caption{The zero-crossing point of the $v_0(\pT)$ obtained for the three $\pTref$ ranges. They are compared to the $\lr{[\pT]}$ calculated for charged particles over 0.5--10~GeV. The data are shifted slightly along the $x$-axis for better visibility. Bars and shaded areas indicate statistical and total uncertainties, respectively.}
\label{fig:app2}
\end{figure}

\begin{figure}[h!]
\centering
\includegraphics[width=0.96\linewidth]{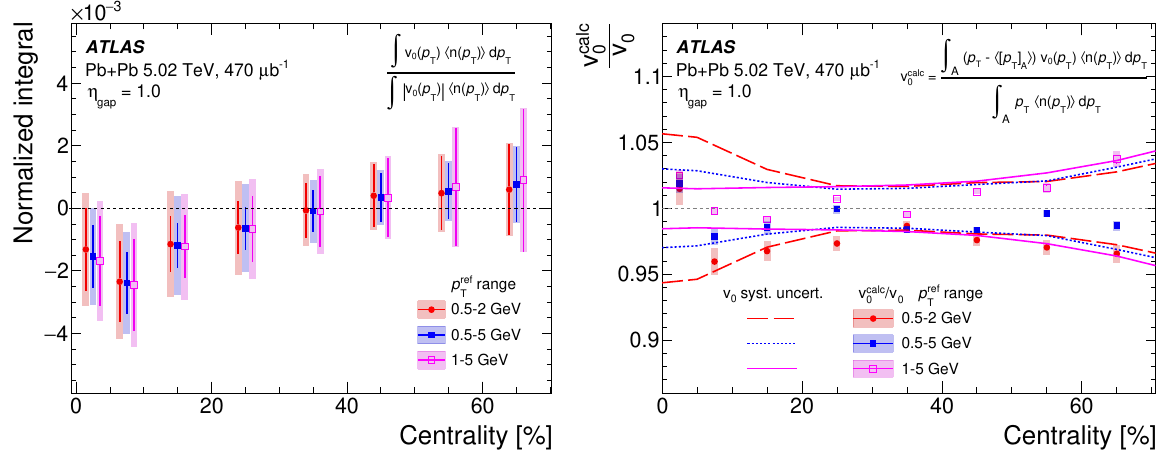}
\caption{The closure of the two sum rules:  $\int v_0(\pT)\lr{n(\pT)} d\pT=0$ (left) and Eq.~\eqref{eq:2} (right) obtained for the three $\pTref$ ranges. Bars and shaded areas indicate statistical and total uncertainties, respectively.}
\label{fig:app3}
\end{figure}

Figure~\ref{fig:app2} shows the zero-crossing point of the $v_0(\pT)$ obtained for the three $\pTref$ ranges, and compared to the $\lr{[\pT]}$ over 0.5--10~GeV. Figure~\ref{fig:app3} shows the performance of the two sum rules of the $v_0(\pT)$ expected due to its collective nature.

\begin{figure}[h!]
\centering
\includegraphics[width=0.8\linewidth]{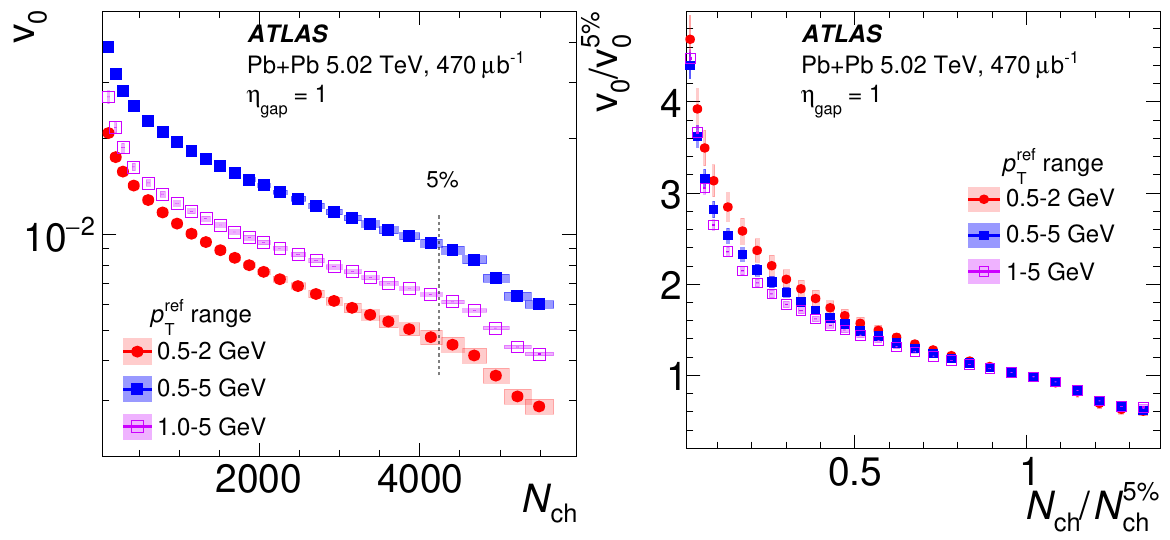}
\caption{Left: $v_0$ as a function of $\nch$ for three $\pTref$ ranges and $\eta_{\mathrm{gap}}=1$. Right: $v_0$ divided by its value at 5\% centrality. Bars and shaded areas indicate statistical and total uncertainties, respectively.}
\label{fig:app4}
\end{figure}

\begin{figure}[h!]
\centering
\includegraphics[width=0.8\linewidth]{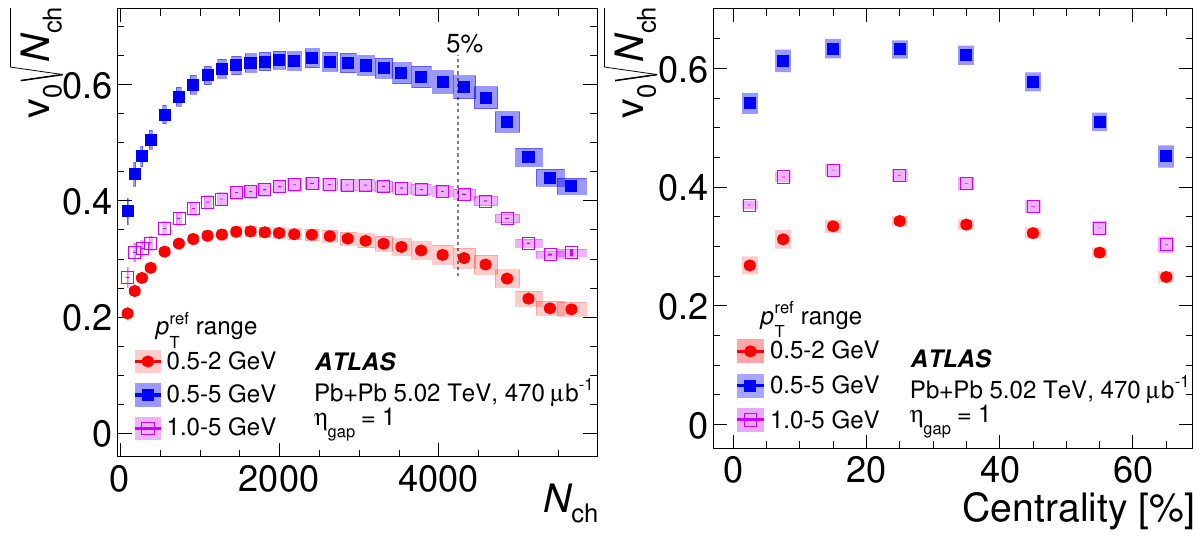}
\caption{$\sqrt{\nch} v_0(\pT)$ as a function of $\nch$ (left) and centrality (right) for three $\pTref$ ranges and $\eta_{\mathrm{gap}}=1$. Bars and shaded areas indicate statistical and total uncertainties, respectively.}
\label{fig:app5}
\end{figure}

\begin{figure}[h!]
\centering
\includegraphics[width=0.7\linewidth]{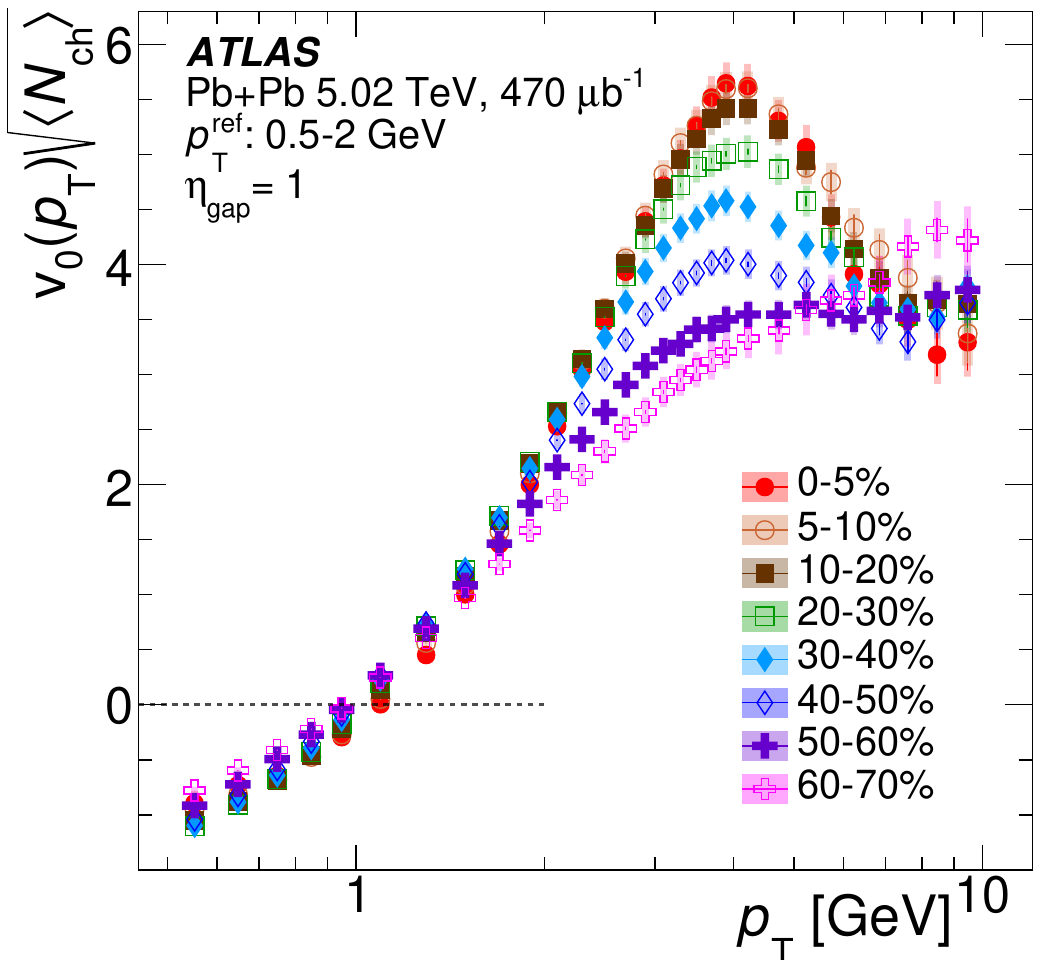}
\caption{The $\sqrt{\nch}v_0(\pT)$ for different centrality ranges. Bars and shaded areas represent statistical and total uncertainties, respectively. }
\label{fig:app6}
\end{figure}

The results for $v_0$ were previously presented as a function of the efficiency-corrected charged-particle multiplicity, $\nch$~\cite{ATLAS:2024jvf}. Figure~\ref{fig:app4} shows the $v_0$ and $v_0$ divided by its value at 5\% as a function of $\nch$, complementing Figure~\ref{fig:2}. The $v_0$ were observed to scale as $\sqrt{\nch}$ in a previous ATLAS paper~\cite{ATLAS:2024jvf}; such scaling is shown in Figure~\ref{fig:app5} for the three $\pTref$ ranges. Figure~\ref{fig:app6} shows that the $v_0(\pT)$ scaled by $\sqrt{\nch}$ approximately falls on a common trend at low $\pT$, expected from the independent source model scenario.


\clearpage
\printbibliography
\clearpage
 
\begin{flushleft}
\hypersetup{urlcolor=black}
{\Large The ATLAS Collaboration}

\bigskip

\AtlasOrcid[0000-0002-6665-4934]{G.~Aad}$^\textrm{\scriptsize 104}$,
\AtlasOrcid[0000-0001-7616-1554]{E.~Aakvaag}$^\textrm{\scriptsize 17}$,
\AtlasOrcid[0000-0002-5888-2734]{B.~Abbott}$^\textrm{\scriptsize 123}$,
\AtlasOrcid[0000-0002-0287-5869]{S.~Abdelhameed}$^\textrm{\scriptsize 119a}$,
\AtlasOrcid[0000-0002-1002-1652]{K.~Abeling}$^\textrm{\scriptsize 55}$,
\AtlasOrcid[0000-0001-5763-2760]{N.J.~Abicht}$^\textrm{\scriptsize 49}$,
\AtlasOrcid[0000-0002-8496-9294]{S.H.~Abidi}$^\textrm{\scriptsize 30}$,
\AtlasOrcid[0009-0003-6578-220X]{M.~Aboelela}$^\textrm{\scriptsize 45}$,
\AtlasOrcid[0000-0002-9987-2292]{A.~Aboulhorma}$^\textrm{\scriptsize 36e}$,
\AtlasOrcid[0000-0001-5329-6640]{H.~Abramowicz}$^\textrm{\scriptsize 157}$
% ...
\bigskip
\\

% ...
$^{17}$Department for Physics and Technology, University of Bergen, Bergen; Norway.\\
% ...
$^{30}$Physics Department, Brookhaven National Laboratory, Upton NY; United States of America.\\
% ...
$^{36}$$^{(a)}$Facult\'e des Sciences Ain Chock, Universit\'e Hassan II de Casablanca;$^{(b)}$Facult\'{e} des Sciences, Universit\'{e} Ibn-Tofail, K\'{e}nitra;$^{(c)}$Facult\'e des Sciences Semlalia, Universit\'e Cadi Ayyad, LPHEA-Marrakech;$^{(d)}$LPMR, Facult\'e des Sciences, Universit\'e Mohamed Premier, Oujda;$^{(e)}$Facult\'e des sciences, Universit\'e Mohammed V, Rabat;$^{(f)}$Institute of Applied Physics, Mohammed VI Polytechnic University, Ben Guerir; Morocco.\\
% ...
$^{45}$Physics Department, Southern Methodist University, Dallas TX; United States of America.\\
% ...
$^{49}$Fakult\"{a}t Physik, Technische Universit{\"a}t Dortmund, Dortmund; Germany.\\
% ...
$^{55}$II. Physikalisches Institut, Georg-August-Universit\"{a}t G\"ottingen, G\"ottingen; Germany.\\
% ...
$^{104}$CPPM, Aix-Marseille Universit\'e, CNRS/IN2P3, Marseille; France.\\
% ...
$^{119}$$^{(a)}$New York University Abu Dhabi, Abu Dhabi;$^{(b)}$United Arab Emirates University, Al Ain; United Arab Emirates.\\
% ...
$^{123}$Homer L. Dodge Department of Physics and Astronomy, University of Oklahoma, Norman OK; United States of America.\\
% ...
$^{157}$Raymond and Beverly Sackler School of Physics and Astronomy, Tel Aviv University, Tel Aviv; Israel.\\
% ...
$^{a}$ Also at Affiliated with an institute formerly covered by a cooperation agreement with CERN.\\
$^{b}$ Also at An-Najah National University, Nablus; Palestine.\\
$^{c}$ Also at Borough of Manhattan Community College, City University of New York, New York NY; United States of America.\\
$^{d}$ Also at Center for Interdisciplinary Research and Innovation (CIRI-AUTH), Thessaloniki; Greece.\\
$^{e}$ Also at Centre of Physics of the Universities of Minho and Porto (CF-UM-UP); Portugal.\\
$^{f}$ Also at CERN, Geneva; Switzerland.\\
% ...

\end{flushleft}


\end{document}